\documentclass[useAMS,usenatbib]{mn2e}

\usepackage{graphicx}
\usepackage{natbib}

\newcommand{\ltsima} {$\; \buildrel < \over \sim \;$}
\newcommand{\gtsima} {$\; \buildrel > \over \sim \;$}
\newcommand{\lta} {\lower.5ex\hbox{\ltsima}}
\newcommand{\gta} {\lower.5ex\hbox{\gtsima}}

\title[Constraints on general primordial non-Gaussianity]{Constraints on general primordial non-Gaussianity using wavelets for the Wilkinson Microwave anisotropy probe 7-year data }

\author[A. Curto et al.]{A. Curto,$^1$ \thanks{e-mail:
curto@ifca.unican.es} E. Mart\'{\i}nez-Gonz\'alez,$^1$ R. B. Barreiro,$^1$ M. P. Hobson$^2$ \\
$^1$     Instituto de F\'isica de Cantabria, CSIC-Universidad de Cantabria, Avda. de los Castros s/n, 39005 Santander, Spain.\\
$^2$Astrophysics Group, Cavendish Laboratory, Madingley Road, Cambridge, CB3 0HE, U.K. \\
}
\date{Accepted  Received ; in original form }

\begin{document}

\maketitle

\begin{abstract}

We present constraints on the non-linear coupling parameter $f_{nl}$
with the Wilkinson Microwave Anisotropy Probe (WMAP) data. We use the
method based on the spherical Mexican hat wavelet (SMHW) to measure
the $f_{nl}$ parameter for three of the most interesting shapes of
primordial non-Gaussianity: {\it local, equilateral} and {\it
orthogonal}. Our results indicate that this parameter is compatible
with a Gaussian distribution within the two sigma confidence level
(CL) for the three shapes and the results are consistent with the
values presented by the WMAP team. We have included in our analysis
the impact on $f_{nl}$ due to contamination by unresolved point
sources. The point sources add a positive contribution of $ \Delta
f^{loc}_{nl}= 2.5 \pm 3.0$, $ \Delta f^{eq}_{nl}= 37 \pm 18$ and
$ \Delta f^{ort}_{nl}= 25 \pm 14$. As mentioned by the WMAP team,
the contribution of the point sources to the orthogonal and
equilateral form is expected to be larger than to the local one and
thus it cannot be neglected in future constraints on these
parameters. Taking into account this contamination, our best estimates
for $f_{nl}$ are $ -16.0\le f^{loc}_{nl} \le 76.0$, $ -382 \le
f^{eq}_{nl} \le 202$ and $ -394 \le f^{ort}_{nl} \le 34$ at 95\%
CL. The three shapes are compatible with zero at 95\% CL
($2\sigma$). Our conclusion is that the WMAP 7-year data are
consistent with Gaussian primordial fluctuations within $2\sigma$ CL.
We stress however the importance of taking into account the unresolved
point sources in the measurement of $f_{nl}$ in future works,
especially when using more precise data sets such as the forthcoming
Planck data.
\end{abstract}
\section{Introduction}

During the period of inflationary expansion in the very early stages
of the universe, primordial perturbations were generated that are the
seeds of the structures that we can observe today
\citep{starobinski1979,guth,albrecht,linde1982,linde1983,mukhanov1992}.
These primordial perturbations were linearly imprinted in the Cosmic
Microwave Background (CMB) anisotropies. Thus the study of the CMB
anisotropies is a powerful way to understand the physics of the early
universe. Many observational CMB projects, for example the NASA
WMAP\footnote{http://map.gsfc.nasa.gov/} and ESA
Planck\footnote{http://www.esa.int/planck} missions, different ground
based 3D observational campaigns of large scale structure and high
energy accelerators are enabling us to understand better the
properties and the evolution of the universe. From the several
observational approaches that are available, the search for departures
from Gaussianity in the CMB anisotropies with a primordial origin has
become a powerful way to discriminate among different inflationary
scenarios. Inflationary models such as the widely accepted standard,
single-field, slow roll inflation predict low levels of
non-Gaussianity whereas other models predict levels of non-Gaussianity
that may be detected using the data from current experiments
\citep{bartolo2004,komatsu2009b,yadav2010,komatsu2010b}. A detection
of a deviation from Gaussianity with a primordial origin would rule
out many inflationary models and would have far reaching implications
in the physics of the early universe.

The level of primordial non-Gaussianity is usually parametrised by the
non-linear coupling parameter $f_{nl}$
\citep{verde2000,komatsu2001,bartolo2004}. This parameter measures
departures from zero in the values of the third order quantity known
as the bispectrum, characterised through the shape function
$F(k_1,k_2,k_3)$. The bispectrum is related to Bardeen's curvature
perturbations $\Phi({\bf k})$ through the 3-point correlation function
$\langle \Phi({\bf k_1}) \Phi({\bf k_2}) \Phi({\bf k_3}) \rangle =
(2\pi)^3\delta^3({\bf k_1+k_2+k_3})F(k_1,k_2,k_3)$. Depending on the
physical mechanisms of the different inflationary models the shape
function can take different forms.

In this paper we measure the levels of non-Gaussianity present in the
WMAP data corresponding to the three particular shapes (local,
equilateral and orthogonal) that have been studied by the WMAP team
\citep{komatsu2011}. The shape function $F(k_1,k_2,k_3)$ of these
types of non-Gaussianity, their CMB angular bispectra $b_{\ell_1
\ell_2 \ell_3}$ and the inflationary scenarios that generate these
non-Gaussianity are described below.
\begin{itemize}
\item {\bf Local shape.} Significant non-Gaussianity of the local form
can be generated for example in multi-field inflationary models
\citep{komatsu2005, komatsu2010b}, the curvaton model
\citep{lyth2003}, the inhomogeneous reheating scenario
\citep{dvali2004,bartolo2004}, models based on hybrid inflation
\citep{lin2009}, etc. This shape is given by \citep[see for
example][]{creminelli2006,fergusson2010a,yadav2010,komatsu2011}
\begin{eqnarray}
\nonumber
F(k_1,k_2,k_3)=2A^2f_{nl}\Big[\frac{1}{k_{1}^{3-(n_s-1)}k_{2}^{3-(n_s-1)}}+\\
+ \frac{1}{k_{1}^{3-(n_s-1)}k_{3}^{3-(n_s-1)}} + \frac{1}{k_{2}^{3-(n_s-1)}k_{3}^{3-(n_s-1)}}\Big],
\end{eqnarray}
and its angular bispectrum is \citep[see for
example][]{fergusson2010a,yadav2010,komatsu2010b}
\begin{eqnarray}
\nonumber
b^{loc}_{\ell_1 \ell_2 \ell_3} = 2\int_{0}^{\infty}x^2dx\Big[\alpha_{\ell_1}(x)\beta_{\ell_2}(x)\beta_{\ell_3}(x) +\\
\beta_{\ell_1}(x)\alpha_{\ell_2}(x)\beta_{\ell_3}(x) + \beta_{\ell_1}(x)\beta_{\ell_2}(x)\alpha_{\ell_3}(x)\Big],
\label{localbispectrum}
\end{eqnarray}
where $A$ is the amplitude of the power spectrum
$P_{\Phi}(k)=Ak^{n_s-4}$, $n_s$ is the spectral index and
$\alpha_{\ell}(x)$, $\beta_{\ell}(x)$ are filter functions \citep[see
  for example][]{komatsu2001,komatsu2005,fergusson2010a,komatsu2010b}.
\item {\bf Equilateral shape.}  Significant non-Gaussianity of the
  equilateral form can be generated for example by the
  Dirac-Born-Infeld inflation
  \citep{silverstein2004,bartolo2004,langlois2008}, ghost inflation
  \citep{arkani-hamed2004}, several single-field inflationary models
  in Einstein gravity \citep{chen2007} etc. This shape is given by
  \citep[see for
    example][]{creminelli2006,fergusson2010a,yadav2010,komatsu2011}
\begin{eqnarray}
\nonumber
F(k_1,k_2,k_3)=6A^2f_{nl}\Big[-\frac{1}{k_{1}^{3-(n_s-1)}k_{2}^{3-(n_s-1)}}\\
\nonumber
- \frac{1}{k_{1}^{3-(n_s-1)}k_{3}^{3-(n_s-1)}} - \frac{1}{k_{2}^{3-(n_s-1)}k_{3}^{3-(n_s-1)}}\\
\nonumber
-\frac{2}{(k_1k_2k_3)^{2(4-n_s)/3}} \\
+ \Big\{\frac{1}{k_1^{(4-n_s)/3}k_2^{2(4-n_s)/3}k_3^{(4-n_s)}} + (5~perm)\Big\}\Big],
\end{eqnarray}
and its angular bispectrum is \citep[see for
example][]{fergusson2010a,yadav2010,komatsu2010b}
\begin{eqnarray}
\nonumber
b^{eq}_{\ell_1 \ell_2 \ell_3} = 6 \int_{0}^{\infty}dxx^2\Big [ -\alpha_{\ell_1}(x)\beta_{\ell_2}(x)\beta_{\ell_3}(x) + (2~perm)\\
 + \beta_{\ell_1}(x)\gamma_{\ell_2}(x)\delta_{\ell_3}(x)  + (5~perm) -2 \delta_{\ell_1}(x)\delta_{\ell_2}(x)\delta_{\ell_3}(x) \Big ],
\label{equilateralbispectrum}
\end{eqnarray}
where $\gamma_{\ell}(x)$ and $\delta_{\ell}(x)$ are filter functions
\citep[see for example][]{fergusson2010a,komatsu2010b}.
\item {\bf Orthogonal shape.} Significant non-Gaussianity of the
orthogonal form can be generated in general single-field models
\citep{cheung2008,senatore2010}. This shape is given by \citep[see for
example][]{senatore2010,yadav2010,komatsu2011}
\begin{eqnarray}
\nonumber
F(k_1,k_2,k_3)=6A^2f_{nl}\Big[-\frac{3}{k_{1}^{3-(n_s-1)}k_{2}^{3-(n_s-1)}}\\
\nonumber
- \frac{3}{k_{1}^{3-(n_s-1)}k_{3}^{3-(n_s-1)}} - \frac{3}{k_{2}^{3-(n_s-1)}k_{3}^{3-(n_s-1)}}\\
\nonumber
-\frac{8}{(k_1k_2k_3)^{2(4-n_s)/3}} \\
 +\Big\{\frac{3}{k_1^{(4-n_s)/3}k_2^{2(4-n_s)/3}k_3^{(4-n_s)}} + (5~perm)\Big\}\Big],
\end{eqnarray}
and its angular bispectrum is \citep[see for
example][]{yadav2010,komatsu2010b}
\begin{eqnarray}
\nonumber
b^{ort}_{\ell_1 \ell_2 \ell_3} = 18 \int_{0}^{\infty}dxx^2\Big [ -\alpha_{\ell_1}(x)\beta_{\ell_2}(x)\beta_{\ell_3}(x) + (2~perm) \\
+ \beta_{\ell_1}(x)\gamma_{\ell_2}(x)\delta_{\ell_3}(x) + (5~perm) -\frac{8}{3} \delta_{\ell_1}(x)\delta_{\ell_2}(x)\delta_{\ell_3}(x) \Big ].
\label{orthogonalbispectrum}
\end{eqnarray}
\end{itemize}
Many studies have been performed to constrain $f_{nl}$, especially for
the local and the equilateral cases. The first constraints on $f_{nl}$
were imposed using data sets with low resolution or small sky coverage
which led to large uncertainties in $f_{nl}$. We can report analyses
using the {\it Cosmic Background Explorer} (COBE) data
\citep{komatsu2002,cayon2003}, MAXIMA data
\citep{cayon2003b,santos2003}, the {\it Very Small Array} (VSA) data
\citep{smith2004}, the Archeops data \citep{curto2007,curto2008} and
the BOOMERang data \citep{troia,natoli2010}.

Once the WMAP data were available, significant improvements were
achieved in the precision of the estimation of $f_{nl}$\footnote{The
  improvement comes from a combination of large sky coverage, high
  angular resolution and good sensitivity. This combination improves
  the signal-to-noise ratio of $f_{nl}$ which for the local case is
  proportional to $log(\ell_{max})$ \citep{yadav2010}.}. Many studies
have been developed to constrain the $f_{nl}$ using WMAP data and
based on different estimators. We can mention the different
bispectrum-based estimators \citep[see for
  example][]{komatsu2003,babich2004,fergusson2007,spergel2007,creminelli2006,creminelli2007,wandelt2008,fergusson2009,komatsu2009,smith2009,elsner2009,bucher2010,liguori2010,senatore2010,smidt2010,fergusson2010a,fergusson2010b,komatsu2011,fergusson2011}. The
bispectrum is the most natural way to constrain $f_{nl}$ given its
linear dependence and the fact that in certain ideal conditions
bispectrum-based estimators may be the optimal way to measure
$f_{nl}$. However, given that the data are contaminated by different
non-Gaussian parasite signals and in most cases only a fraction of the
sky can be used, it is convenient to use additional tools that can
help to understand these effects better. We can mention the tests
performed using the spherical Mexican hat wavelet (SMHW)
\citep{mukherjee2004,curto2009a,curto2009b,curto2011}, a HEALPix-based
wavelet \citep{casaponsa2011a,casaponsa2011b}, a joint analysis with
the SMHW and neural networks \citep{casaponsa2011b}, needlets
\citep{marinucci2008,pietrobon2009,rudjord2009,pietrobon2010,rudjord2010,pietrobon2010b,pietrobon2010c,cabella2010},
the Minkowski functionals
\citep{hikage2006,gott2007,hikage2008,matsubara2010,takeuchi2010}, the
N-PDF distribution \citep{vielva2009,vielva2010} or a Bayesian
approach \citep{elsner2010a,elsner2010b}. Other works use the 3D
distribution of matter on large scales \citep[see for
  example][]{dalal2008,matarrese2008,slosar2008,seljak2009,desjacques2010,xia2010,baldauf2011,hamaus2011}
to constrain the local $f_{nl}$.

In this paper we focus on the measurement of non-Gaussianity for the
previous mentioned shapes using the estimator based on wavelets that
has been formerly used to constrain local $f_{nl}$
\citep{curto2009a,curto2009b,curto2010,curto2011}. We use the
technique described by \citet{fergusson2010a} to produce non-Gaussian
maps with the local, equilateral and orthogonal bispectra for WMAP
resolution in realistic conditions of partial sky coverage and
anisotropic noise. These maps are later used to evaluate the expected
values of the wavelet third order moments $\alpha_{ijk}$ for each type
of non-Gaussianity. We finally impose constrains on $f_{nl}$ for each
shape using the wavelet estimator for the WMAP foreground reduced and
raw data maps. As shown later, unresolved point sources produce a
significant bias in $f_{nl}$ that should be considered in the analyses
of WMAP data and in the forthcoming analyses of Planck data,
especially for the equilateral and orthogonal shapes.

This paper is organized as follows. Section \ref{nongaussmaps}
presents the non-Gaussian maps that we have used to estimate the
quantities needed for this analysis. In Section \ref{methodsec} we
present the method and the estimator used in this analysis to
constrain $f_{nl}$. The results of the analysis using WMAP data are
presented in Section \ref{applicationwmap} and the conclusions are
presented in Section \ref{conclusions}.
\section{Non-Gaussian simulations}
\label{nongaussmaps}
\begin{figure*}
  \center
  \includegraphics[height=3.4cm,width=5.5cm]{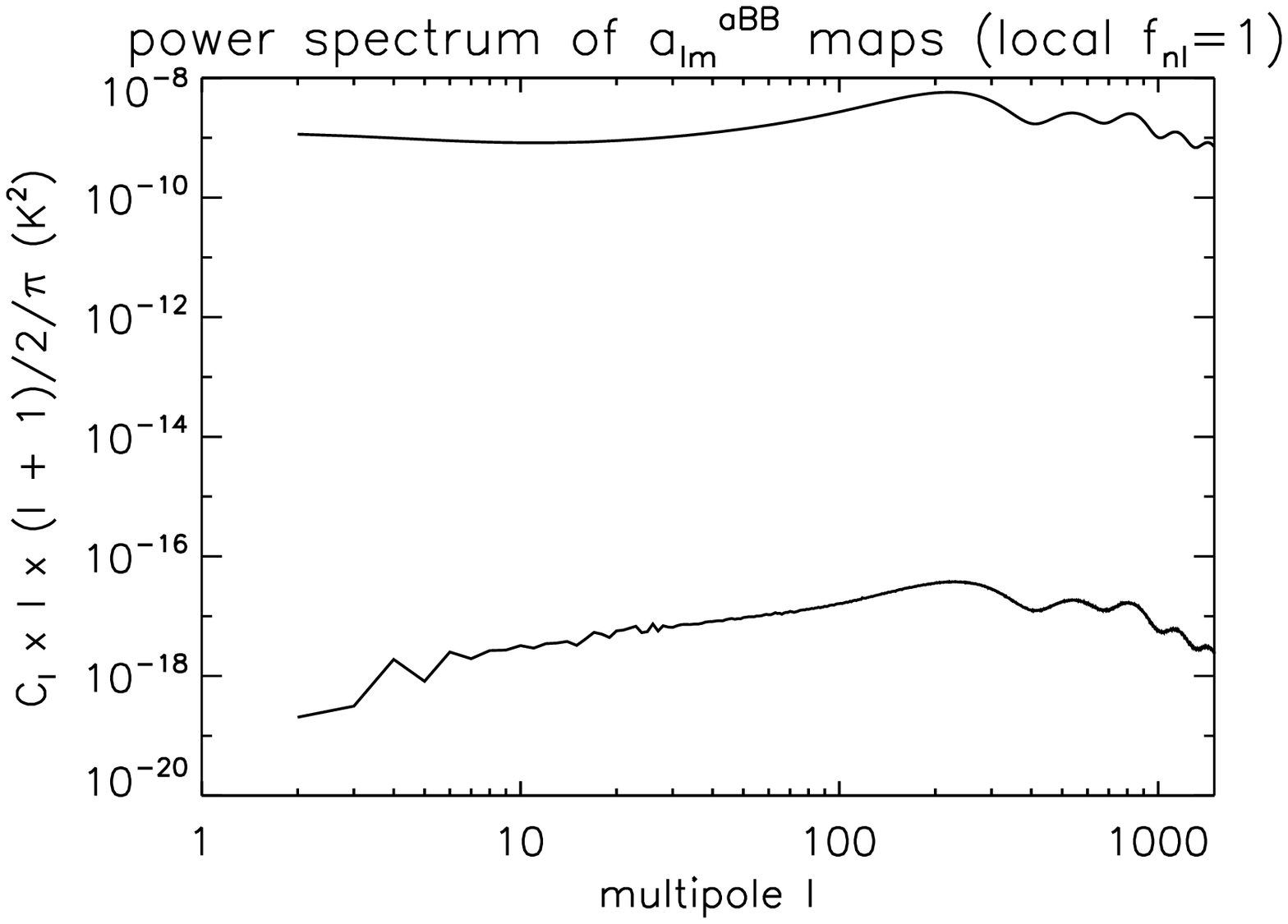}
  \includegraphics[height=3.4cm,width=5.5cm]{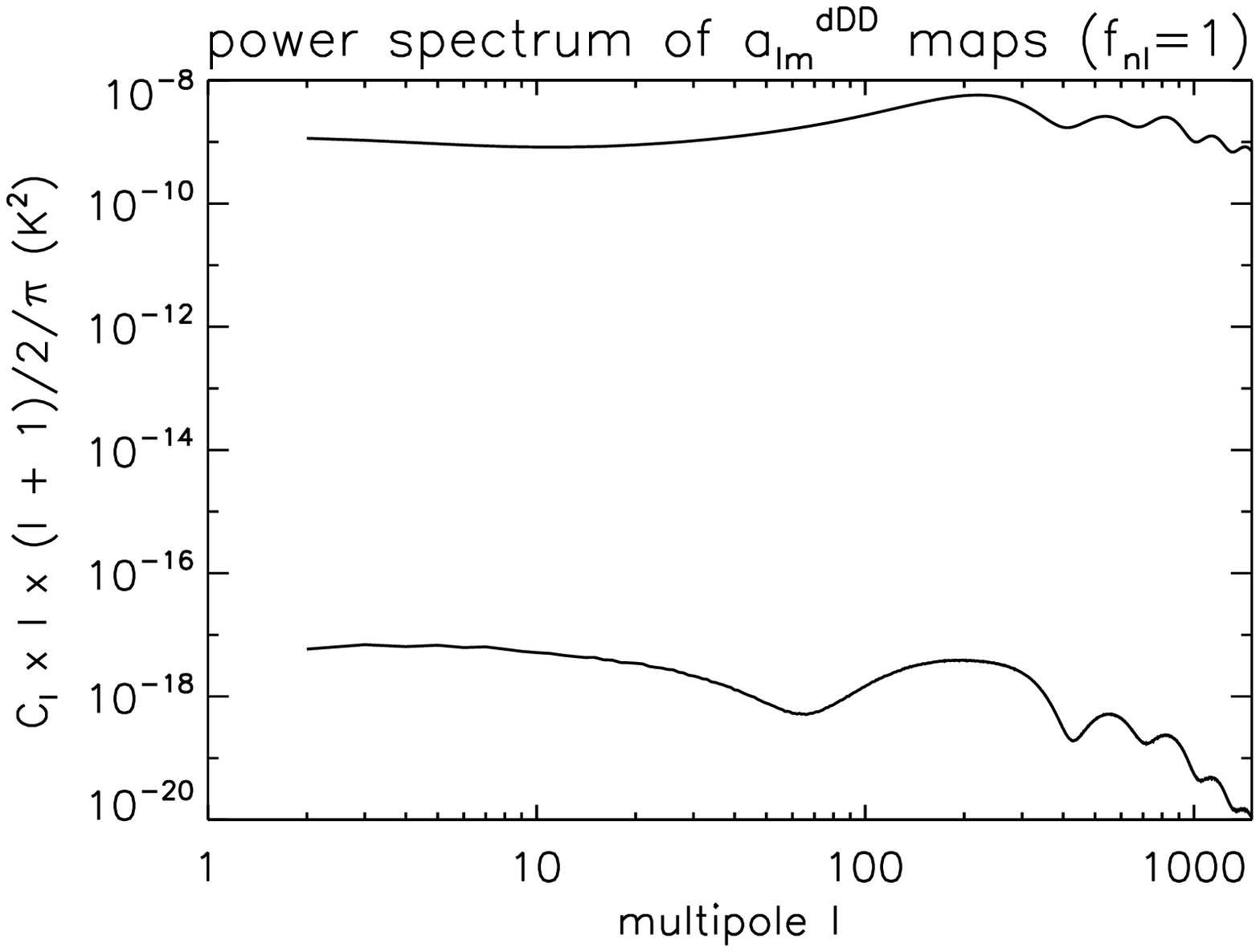}
  \includegraphics[height=3.4cm,width=5.5cm]{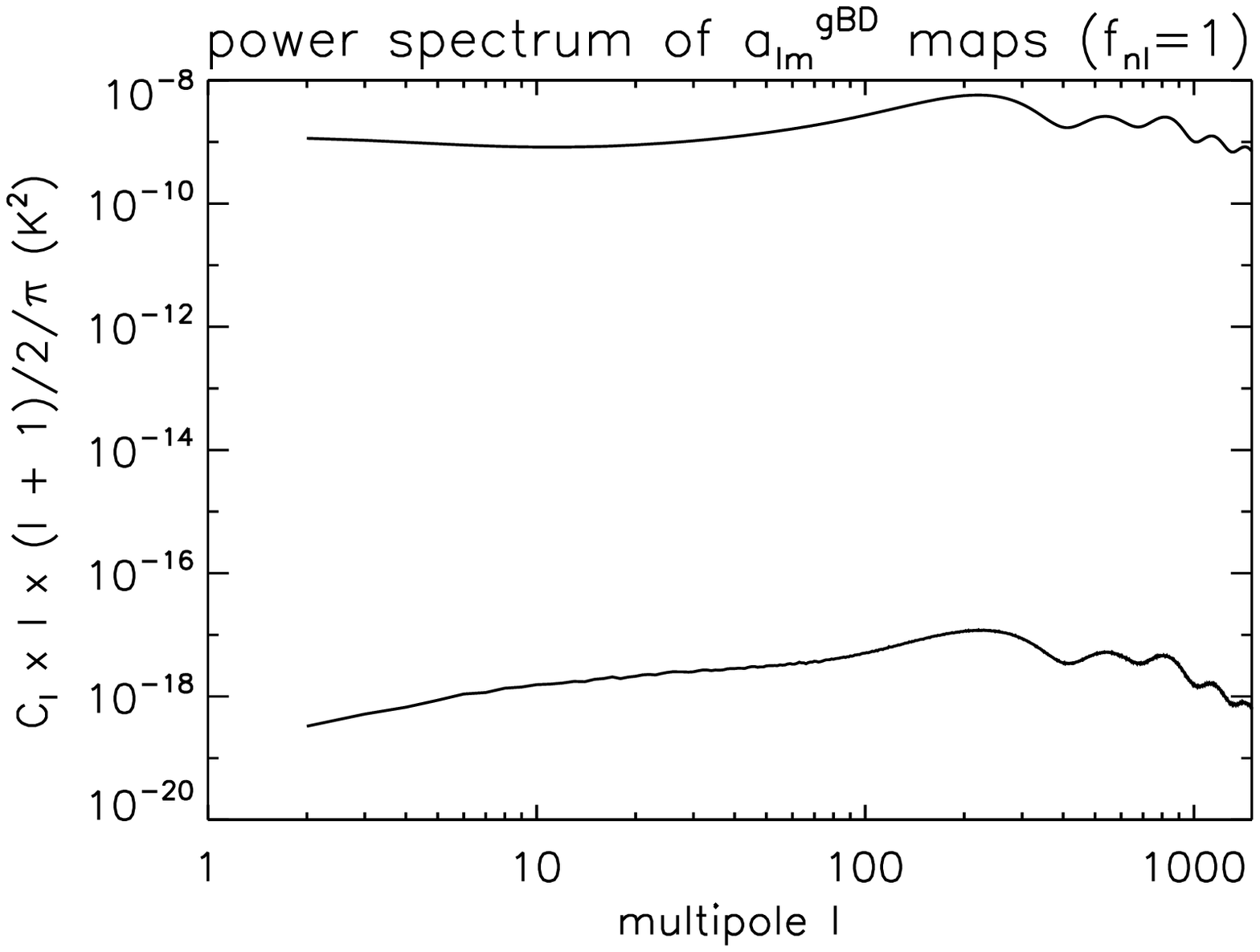}
  \caption{From left to right, the power spectrum of the non-Gaussian
  $a_{\ell m}^{(aBB)}$, $a_{\ell m}^{(dDD)}$ and $a_{\ell m}^{(gBD)}$
  coefficients (lower line) compared with the Gaussian part of the
  power spectrum (upper line). \label{plot_cl_ng_part}}
\end{figure*}

Non-Gaussian Monte Carlo simulations are needed in order to calibrate
the wavelet estimator. We have simulated our non-Gaussian maps
following the algorithm described by \citet{fergusson2010a}. The
non-Gaussian $a^{NG}_{\ell m}$ coefficients can be written in terms of
the bispectrum and the Gaussian $a^{G}_{\ell m}$ coefficients:
\begin{eqnarray}
\nonumber a^{NG}_{\ell m} = \frac{1}{6}\sum_{\ell_2,m_2,\ell_3,m_3}
b_{\ell\ell_2\ell_3}G_{\ell\ell_2\ell_3}^{m m_2m_3} \\
\times \left ( \begin{array}{ccc} \ell & \ell_2 & \ell_3 \\ m & m_2 & m_3
  \end{array}\right ) 
 \frac{a_{\ell_2
  m_2}^{G*}}{C_{\ell_2}}\frac{a_{\ell_3 m_3}^{G*}}{C_{\ell_3}}.
\label{nongaussalm}
\end{eqnarray}
Using the fact that the shape functions of the local, equilateral and
orthogonal bispectra are separable, we are able to reduce the number
of sums in Eq. \ref{nongaussalm}. This can be done in a
straightforward way using Eqs. \ref{localbispectrum},
\ref{equilateralbispectrum} and \ref{orthogonalbispectrum} in
Eq. \ref{nongaussalm}. However, as stated by
\citet{hanson2009,fergusson2010a}, there are terms that may produce
spurious divergences at low multipoles, large enough to affect the
power spectrum of the final map. \citet{fergusson2010a} located the
divergent terms and provided equations for the local and equilateral
shapes without these terms. A similar procedure can be performed with
the orthogonal shape. In the next equations we present the
non-Gaussian $a^{NG}_{\ell m}$ coefficients for each of the shapes
without divergent terms.
\begin{itemize}
\item Local bispectrum
\end{itemize}
\begin{equation}
a_{\ell m}^{NG} = \int_{0}^{\infty}dxx^2\alpha_{\ell}(x)\int d^2\vec{n}Y_{\ell m}^{*}(\vec{n})M_{\beta}(x,\vec{n})M_{\beta}(x,\vec{n})
\label{alm_local}
\end{equation}
\begin{itemize}
\item Equilateral bispectrum
\end{itemize}
\begin{eqnarray}
\nonumber
a_{\ell m}^{NG} = \int_{0}^{\infty}dxx^2\Big \{ -3 \alpha_{\ell}(x)\int d^2\vec{n}Y_{\ell m}^{*}(\vec{n})M_{\beta}(x,\vec{n})M_{\beta}(x,\vec{n}) \\
\nonumber
-2 \delta_{\ell}(x)\int d^2\vec{n}Y_{\ell m}^{*}(\vec{n})M_{\delta}(x,\vec{n})M_{\delta}(x,\vec{n}) \\
+6 \gamma_{\ell}(x)\int d^2\vec{n}Y_{\ell m}^{*}(\vec{n})M_{\beta}(x,\vec{n})M_{\delta}(x,\vec{n}) \Big \}
\label{alm_equilat}
\end{eqnarray}
\begin{itemize}
\item Orthogonal bispectrum
\end{itemize}
\begin{eqnarray}
\nonumber
a_{\ell m}^{NG} = \int_{0}^{\infty}dxx^2\Big \{ -9 \alpha_{\ell}(x)\int d^2\vec{n}Y_{\ell m}^{*}(\vec{n})M_{\beta}(x,\vec{n})M_{\beta}(x,\vec{n}) \\
\nonumber
-8 \delta_{\ell}(x)\int d^2\vec{n}Y_{\ell m}^{*}(\vec{n})M_{\delta}(x,\vec{n})M_{\delta}(x,\vec{n}) \\
+18 \gamma_{\ell}(x)\int d^2\vec{n}Y_{\ell m}^{*}(\vec{n})M_{\beta}(x,\vec{n})M_{\delta}(x,\vec{n}) \Big \}
\label{alm_ortho}
\end{eqnarray}
where 
\begin{eqnarray}
\alpha_{\ell}(x)&=&\frac{2}{\pi}\int_0^{\infty}k^2dkg_{T\ell}(k)j_{\ell}(kx)\\
\beta_{\ell}(x)&=&\frac{2}{\pi}\int_0^{\infty}k^2dkP_{\Phi}(k)g_{T\ell}(k)j_{\ell}(kx)\\
\gamma_{\ell}(x)&=&\frac{2}{\pi}\int_0^{\infty}k^2dkP^{1/3}_{\Phi}(k)g_{T\ell}(k)j_{\ell}(kx)\\
\delta_{\ell}(x)&=&\frac{2}{\pi}\int_0^{\infty}k^2dkP^{2/3}_{\Phi}(k)g_{T\ell}(k)j_{\ell}(kx),
\end{eqnarray}
$g_{T\ell}(k)$ is the radiation transfer function that can be
evaluated using for example the CAMB\footnote{http://camb.info/} or
gTFast\footnote{http://gyudon.as.utexas.edu/$\sim$komatsu/CRL/index.html}
software, $P_{\Phi}(k)$ is the linear power spectrum, $j_{\ell}(kx)$
is the spherical Bessel function (of the first kind) and the
$M(x, \hat{n})$ maps are defined as 
\begin{eqnarray}
M_{\alpha}(x, \hat{n}) = \sum_{\ell,m}\alpha_{\ell}(x)a_{\ell,m}^G\frac{Y_{\ell,m}(\hat{n})}{C_{\ell}} \\
M_{\beta}(x, \hat{n}) = \sum_{\ell,m}\beta_{\ell}(x)a_{\ell,m}^G\frac{Y_{\ell,m}(\hat{n})}{C_{\ell}} \\
M_{\gamma}(x, \hat{n}) = \sum_{\ell,m}\gamma_{\ell}(x)a_{\ell,m}^G\frac{Y_{\ell,m}(\hat{n})}{C_{\ell}} \\
M_{\delta}(x, \hat{n}) = \sum_{\ell,m}\delta_{\ell}(x)a_{\ell,m}^G\frac{Y_{\ell,m}(\hat{n})}{C_{\ell}}.
\label{M_alpha_beta_gamma_delta_func}
\end{eqnarray}
In Figure \ref{plot_cl_ng_part} we plot the power spectrum of the
non-Gaussian terms $a_{\ell m}^{(aBB)}\equiv\int_{0}^{\infty}dxx^2
\alpha_{\ell}(x)\int d^2\vec{n}Y_{\ell
m}^{*}(\vec{n})M_{\beta}(x,\vec{n})M_{\beta}(x,\vec{n})$, $a_{\ell
m}^{(dDD)}\equiv\int_{0}^{\infty}dxx^2 \delta_{\ell}(x)\int
d^2\vec{n}Y_{\ell
m}^{*}(\vec{n})M_{\delta}(x,\vec{n})M_{\delta}(x,\vec{n})$ and
$a_{\ell m}^{(gBD)}\equiv\int_{0}^{\infty}dxx^2 \gamma_{\ell}(x)\int
d^2\vec{n}Y_{\ell
m}^{*}(\vec{n})M_{\beta}(x,\vec{n})M_{\delta}(x,\vec{n})$. The power
spectrum of the Gaussian part is also plotted. We can see that these
three terms add negligible extra-power to the full Gaussian plus
non-Gaussian map.
Once the $a_{\ell m}^{NG}$ terms are computed as a function of the
bispectrum and the $a_{\ell m}^{G}$, the $a_{\ell m}$ coefficients of
a simulation with a given $f_{nl}$ can be written as $a_{\ell m} =
a_{\ell m}^{G} + f_{nl} a_{\ell m}^{NG}$.

\begin{figure*}
  \center
  \includegraphics[height=3.4cm,width=5.5cm]{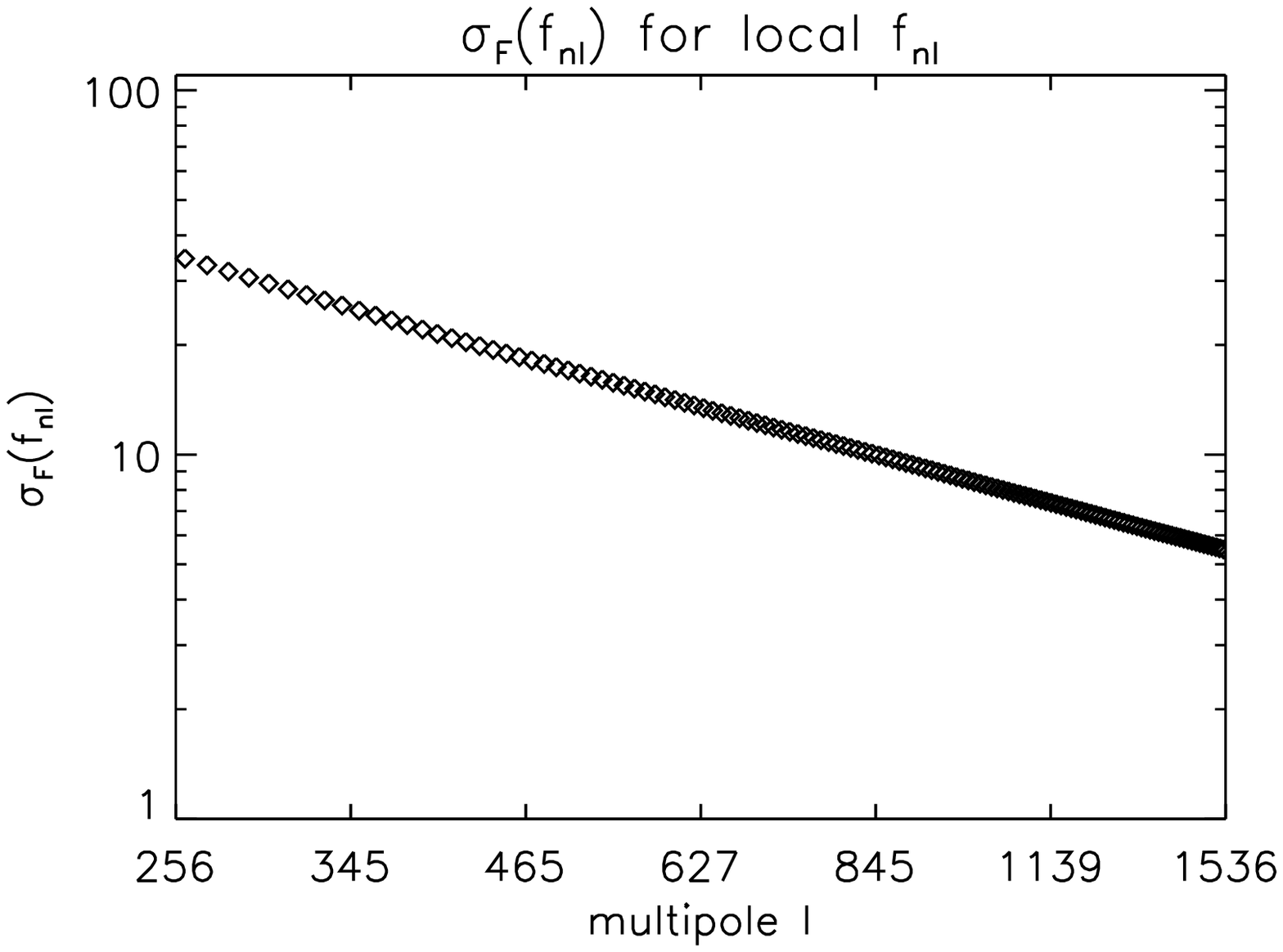}
  \includegraphics[height=3.4cm,width=5.5cm]{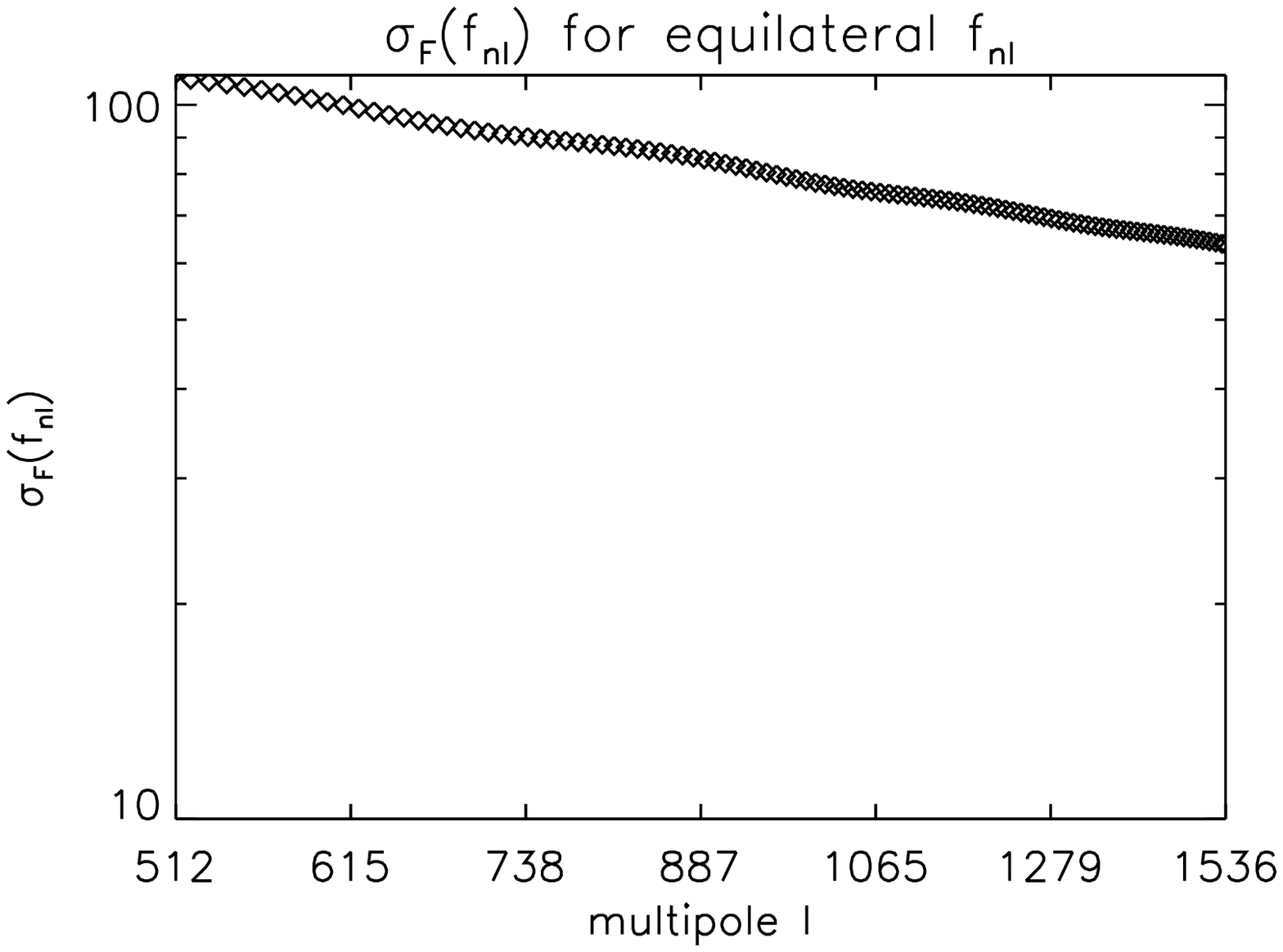}
  \includegraphics[height=3.4cm,width=5.5cm]{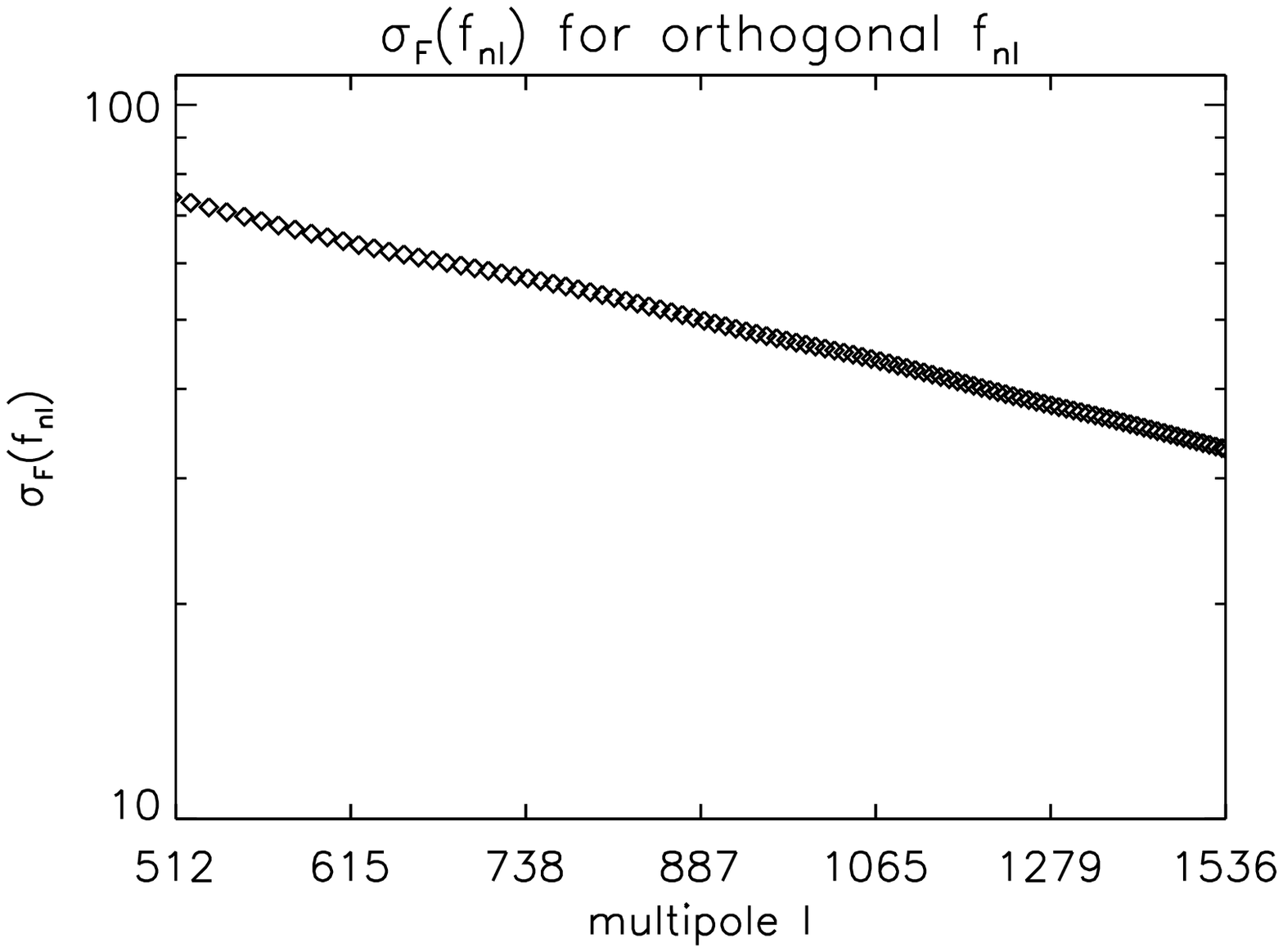}
  \caption{From left to right, the Fisher matrix $\sigma_F(f_{nl})$
  versus $\ell_{max}$ for the local, equilateral and orthogonal
  bispectra $b_{\ell_1\ell_2\ell_3}$ defined in
  Eqs. \ref{localbispectrum}, \ref{equilateralbispectrum} and
  \ref{orthogonalbispectrum}. \label{plot_sigma_fisher}}
\end{figure*}

In this paper we have generated a set of 300 non-Gaussian maps for the
local, equilateral and orthogonal $f_{nl}$. We have assumed a $\Lambda
CDM$ model using the parameters that best fit the WMAP 7-year data
\citep{komatsu2011}. We have computed a power spectrum $C_{\ell}$ and
a transfer function $g_{T \ell}(k)$ using these parameters as inputs
for the CAMB software \citep{lewis2000} up to $\ell_{max}=1535$. The
integrals in Eqs.  \ref{alm_local}, \ref{alm_equilat} and
\ref{alm_ortho} have been performed using a Gauss-Legendre
quadrature. We have used a large density of points near reionization
and recombination (see Table \ref{table_r_quad} for more details). A
large number of points has been chosen in order to achieve convergence
in the values of the Fisher matrix of the
bispectrum. $\sigma_F^2(f_{nl})$ for the three shapes. In Figure
\ref{plot_sigma_fisher} the Fisher matrix $\sigma_F^2(f_{nl})$
\citep{komatsu2001} obtained with the three bispectra is plotted for
different $\ell_{max}$ values. Note that these values are comparable
with the values presented for example by \citet{yadav2010}.
\begin{table}
  \center
  \caption{ Quadrature in x integration used to compute $a_{\ell
m}^{(NG)}$. We have used greater density of points near reionization
and recombination as suggested by \citet{smith2006}. Units for x
are Mpc.
\label{table_r_quad}}
  \begin{tabular}{|c|c|}
    \hline 
    \hline
    $0 \le x \le 9,500$ & 64 points, Gauss-Legendre quadrature\\
    \hline
    $9,500 \le x \le 11,000$ & 128 points, Gauss-Legendre quadrature\\
    \hline
    $11,000 \le x \le 13,800$ & 64 points, Gauss-Legendre quadrature\\
    \hline
    $13,800 \le x \le 14,600$ & 170 points, Gauss-Legendre quadrature\\
    \hline
    $14,600 \le x \le 16,000$ & 42 points, Gauss-Legendre quadrature\\
    \hline
    $16,000 \le x \le 50,000$ & 42 points, Gauss-Legendre quadrature\\
    \hline
    \hline
  \end{tabular}
\end{table}
Once the $a_{\ell m}$ of the simulations with non-Gaussianity are
generated, we transform them into WMAP maps for each radiometer by
convolving with the appropriate window functions in the spherical
harmonic space and by adding a Gaussian instrumental noise simulation
in the real space \citep{bennett2003}.
\section{Method}
\label{methodsec}
We use an estimator that is based on third-order statistics generated
by the different possible combinations of the wavelet coefficient maps
of the SMHW evaluated at certain angular scales. See for example
\citet{antoine1998,martinez2002,vielva2007b,martinez2008} for detailed
information about this wavelet. This estimator is described and used
to search for blind non-Gaussian deviations and constrain local
$f_{nl}$ in \citet{curto2009a,curto2009b,curto2011}. 

We consider the same set of angular scales $R_i$ selected in
\citet{curto2011}. After evaluating the wavelet coefficient map
$w(R_i;{\bf b})$ for each angular scale $R_i$ we compute the third
order moments $q_{ijk}$ for each possible combination of three angular
scales $\{i,j,k\}$.  As mentioned in \citet{curto2011}, the expected
values of the cubic statistics are linearly proportional to $f_{nl}$
\begin{equation}
\langle q_{ijk} \rangle_{f_{nl}} = \alpha_{ijk} f_{nl},
\end{equation}
where the $\alpha_{ijk}$ term is linearly related to the
bispectrum. We evaluate these $\alpha_{ijk}$ quantities for the local,
equilateral and orthogonal bispectra by averaging the values of the
estimators obtained with the non-Gaussian simulations described in the
previous Section. We then compute a $\chi^2$ statistic in order to
constrain each $f_{nl}$
\begin{equation}
\chi^2(f_{nl}) =
\sum_{ijk,rst}(q_{ijk}^{obs}-\alpha_{ijk}f_{nl})C^{-1}_{ijk,rst}(q_{rst}^{obs}-\
\alpha_{rst}f_{nl}),
\label{chi_analytic_fnl}
\end{equation}
where $q_{ijk}^{obs}$ is the value of the statistics obtained for the
actual data map and $C$ is the covariance matrix among the different
statistics $q_{ijk}$\footnote{Note that for this estimator there is no
need to subtract any linear term due to the anisotropic noise as in
the case of the KSW estimator. The reason is that the non-ideal
aspects of the analysis (as the mask, the anisotropic noise, etc.) are
included in the covariance matrix and the $\alpha_{ijk}$
coefficients.}. The covariance matrix is estimated using the $q_{ijk}$
statistics corresponding to 10,000 WMAP Gaussian simulations. A
detailed study described in \citet{curto2011} was carried out in order
to compute correctly its inverse avoiding possible degeneracies. The
$\alpha_{ijk}$ statistics are estimated using the set of 300
non-Gaussian simulations transformed into WMAP $V+W$ maps. Although we
found analytical expressions for the covariance matrix $C_{ijk,rst}$
and the $\alpha_{ijk}$ quantities \citep[see][]{curto2011}, those
expressions are only valid for the particular ideal case of full sky
maps and white isotropic noise. For a realistic case, the analytical
expressions become more complicated and the best practical approach to
compute those quantities is using simulations.

Finally, this estimator is also applied to a set of Gaussian maps in
order to obtain an empirical estimate of the uncertainties of
$f_{nl}$. Additionally we also compute the value of the $f_{nl}$
Fisher matrix using the wavelet coefficients
\begin{equation}
\nonumber \sigma_F^2(f_{nl}) =
\frac{1}{\sum_{ijk,rst}\alpha_{ijk}C^{-1}_{ijk,rst}\alpha_{rst}}.
\label{sigma_fnl_likelihood}
\end{equation}
\section{Application to WMAP data}
\label{applicationwmap}
\subsection{Constraints on $f_{nl}$ using WMAP data}
We use the combined WMAP 7-year V and W band maps at the HEALPix
\citep{healpix} resolution of $N_{side}=512$. We consider both raw and
foreground reduced data maps as \citet{komatsu2011}. The maximum
multipole chosen in this analysis is $3 N_{side}$ although the noise
contamination starts to be significant at $\ell \sim 1000$. For the
three shapes we find the best limits on $f_{nl}$ and provide the value
of the Fisher and the simulated $\sigma(f_{nl})$. During all the
analysis we use the WMAP KQ75 mask \citep{gold2011}. In Table
\ref{table_fnl_results} we summarize our results.
\begin{table}
  \center
  \caption{Constraints on the local, equilateral and orthogonal
  $f_{nl}$ for the clean and raw maps and their uncertainties obtained
  with the Fisher matrix $\sigma_{F}(f_{nl})$ and with simulations
  (RMS).
\label{table_fnl_results}}
  \begin{tabular}{|cccc|cc|}
    \hline \hline 
    CASE & raw $f_{nl}$ & clean $f_{nl}$ &  $\sigma_{F}(f_{nl})$ \ & MEAN & RMS \\
    \hline
    local & 25.0 & 32.5 & 22.5 & 0.0 & 23.00 \\
    equilateral & 28.0 & -53.0 & 145.0 & 1.0 & 156.0 \\
    orthogonal & -119.0 & -155.0 & 106.0 & 0.0 & 112.00 \\
    \hline
    \hline
  \end{tabular}
\end{table}
We find that for the three cases, the parameters are compatible with
zero at 95\% CL. We would like to note the different effect that the
foregrounds produce on different shapes: whereas it is negative for
the local shape, it is positive for the equilateral and orthogonal
shapes. For all the cases, $\sigma_F(f_{nl})$ is lower than the value
obtained with simulations ($\sim$95\% depending on the shape). We
think that this small discrepancy is due to the limited number of
simulations. We have checked that our estimator is unbiased. We have
estimated the $ f_{nl}$ values of 100 non-Gaussian simulations with an
input $ f_{nl}= 100$ and used the remaining 200 non-Gaussian
simulations to estimate the $ \alpha_{ijk}$. The results are $
f^{loc}_{nl} =99.5 \pm 29.5$, $ f^{eq}_{nl} =98 \pm 150$ and $
f^{ort}_{nl} =97 \pm 118$, which are clearly compatible with the input
$ f_{nl}$ taking into account the expected errors in the mean for the
available number of realizations.  Our best estimates for the clean
maps are:
\begin{itemize}
\item Local\footnote{Using a set of 300 non-Gaussian simulations
generated following the procedure by \citet{liguori2003,liguori2007}
our best estimate is $f_{nl}=37 \pm 21$ (68\% CL). These maps were
generated in a different way: the non-Gaussianity is introduced in the
primordial curvature perturbation $\Phi({\bf x})= \Phi_L({\bf x})+
f_{nl}\left(\Phi_L^2({\bf x})-\langle \Phi_L^2({\bf x}) \rangle
\right)$ and then extrapolated to the CMB. This process add extra
non-Gaussianity at higher moments whereas the procedure used in this
paper and in \citet{fergusson2010a} just adds non-Gaussianity to the
third order moments (bispectrum).}: $f_{nl}=32.5 \pm 22.5$ (68\% CL)
\item Equilateral: $f_{nl}=-53 \pm 145$ (68\% CL)
\item Orthogonal: $f_{nl}=-155 \pm 106$ (68\% CL)
\end{itemize}
The values match well the results presented by \citet{komatsu2011}
within one sigma error-bars. The differences can be explained by the
different sensitivity of the bispectrum and wavelet estimators to
the possible non-cosmological residuals present in the data.
\subsection{Point source contribution}
We have also estimated the contribution of undetected point sources
using the source number counts dN/dS derived from \citet{zotti}.  We
have used point source simulations based on this dN/dS. We have chosen
a maximum flux for the bright sources such that the power spectrum for
the Q band is compatible with the value provided by the WMAP team,
$ A_{ps} = 0.0090 \pm 0.0007 ~\mu K^2 ~sr$ in antenna units
\citep{larson2011}. We have estimated the best-fitting $f_{nl}$ value
for two sets of 1,000 maps. The first set consists of 1,000 Gaussian
CMB + noise maps and the second consists of the same Gaussian CMB +
noise maps plus the point source maps. For each map with point sources
we estimate its best-fitting $f_{nl}$ parameter and compare with the
value obtained for the same map without point sources. The difference
$\Delta f_{nl}$ provides an estimate of the impact on $f_{nl}$ due to
the unresolved point sources. The point sources add a contribution of
$ \Delta f_{nl}= 2.5 \pm 3$, $ \Delta f_{nl}= 37 \pm 18$ and
$ \Delta f_{nl}= 25 \pm 14$ for the local, equilateral and
orthogonal forms respectively.

To check further these results, we have used an alternative method to
estimate the point source contamination to $f_{nl}$ given by the
expression
\begin{equation}
\Delta f_{nl} = \frac{\sum_{ijk,rst}\langle q_{ijk} \rangle_{ps}C^{-1}_{ijk,rst}\alpha_{rst}}{\sum_{ijk,rst}\alpha_{ijk}C^{-1}_{ijk,rst}\alpha_{rst}},
\end{equation}
where $ \langle q_{ijk} \rangle_{ps}$ is the expected value of the
third order moments due to the point sources. The results are $ \Delta
f_{nl}= 2.5$, $ \Delta f_{nl}= 38$ and $ \Delta f_{nl}= 24$ which
agree with the values previously obtained with simulations. Taking
into account the point source contribution, our best estimates of
$f_{nl}$ are:
\begin{itemize}
\item Local: $f_{nl}=30.0 \pm 22.5$ (68\% CL)
\item Equilateral: $f_{nl}=-90 \pm 146$ (68\% CL)
\item Orthogonal: $f_{nl}=-180 \pm 107$ (68\% CL)
\end{itemize}
\begin{figure*}
  \center
  \includegraphics[height=3.4cm,width=5.5cm]{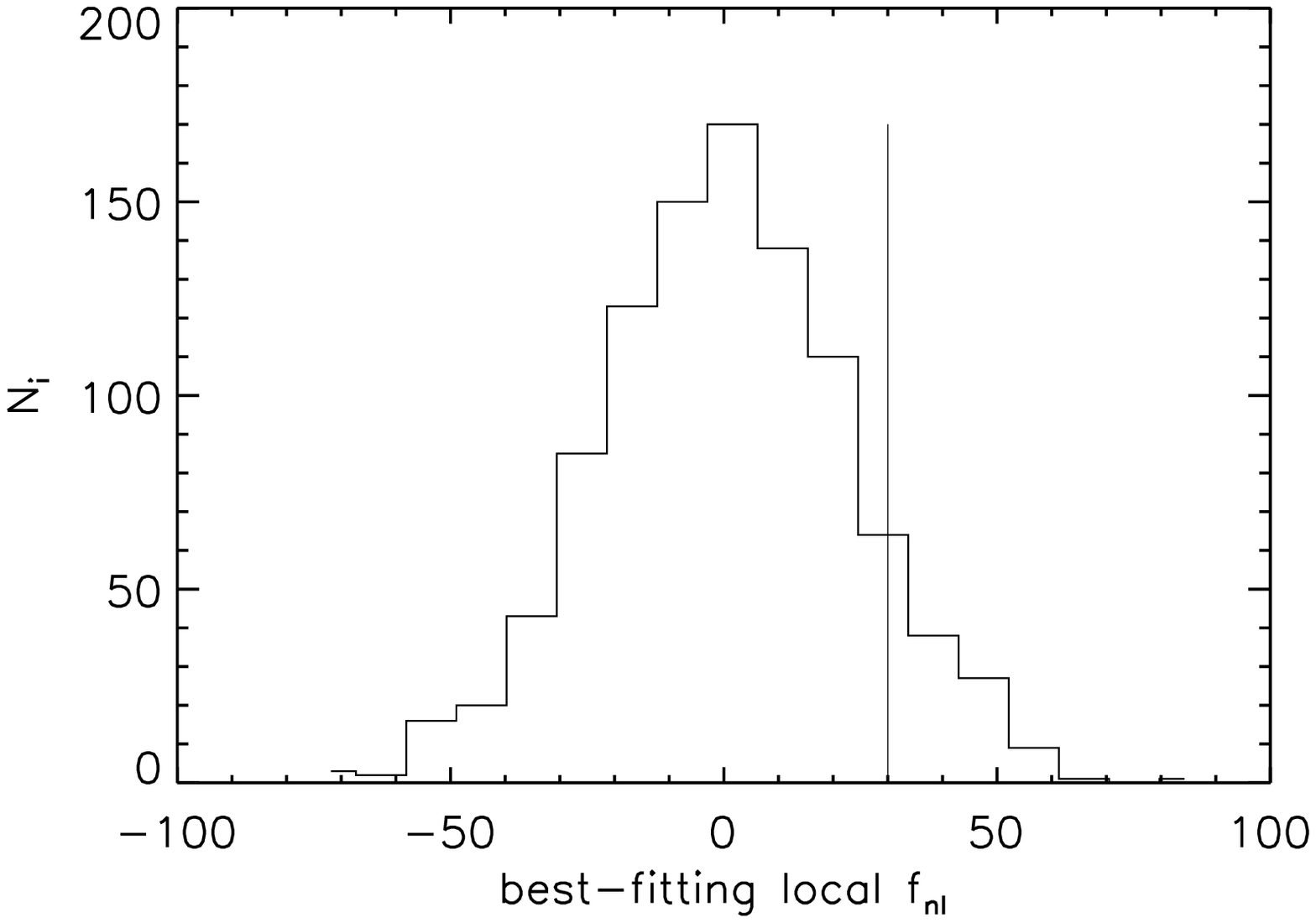}
  \includegraphics[height=3.4cm,width=5.5cm]{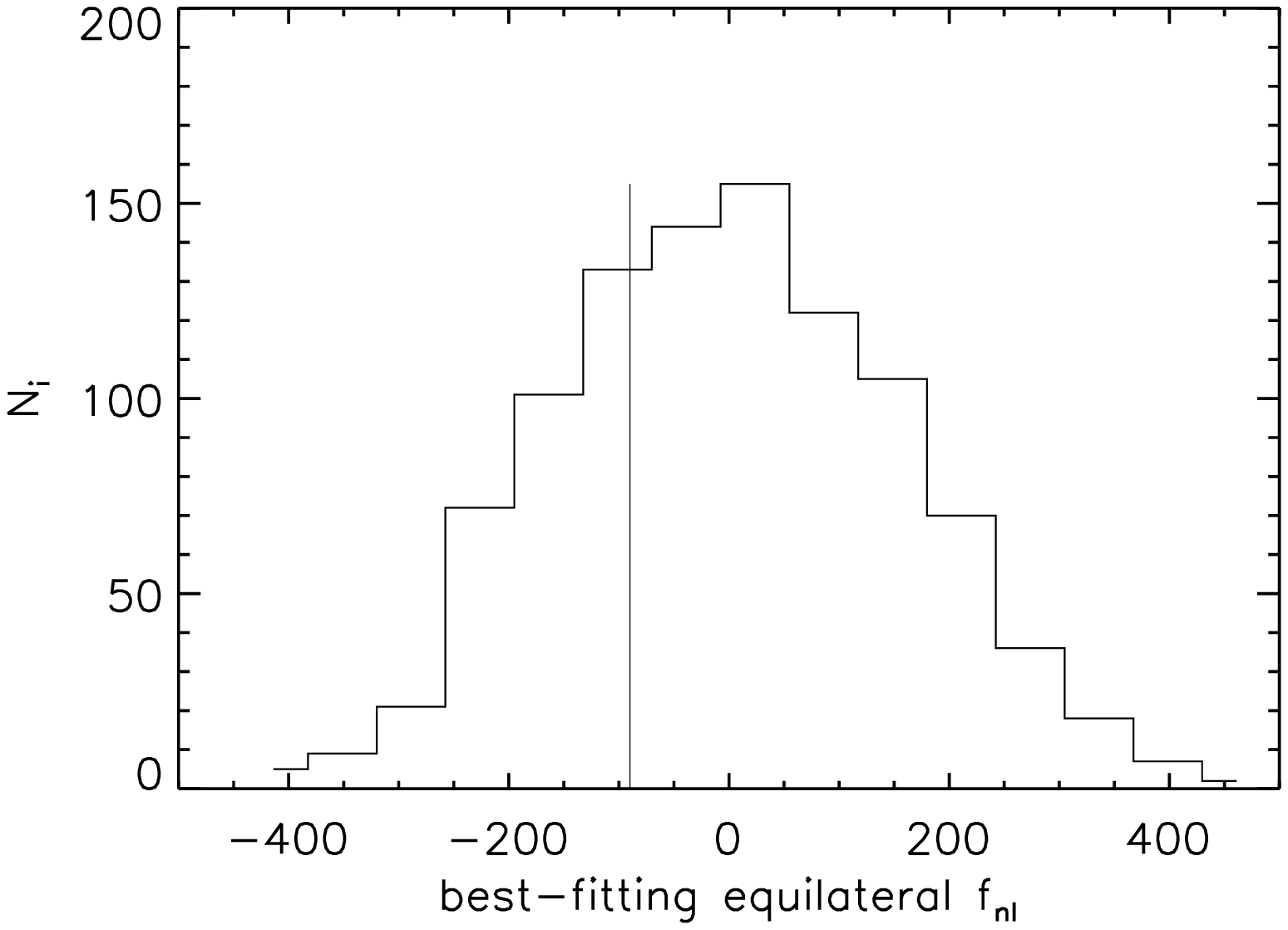}
  \includegraphics[height=3.4cm,width=5.5cm]{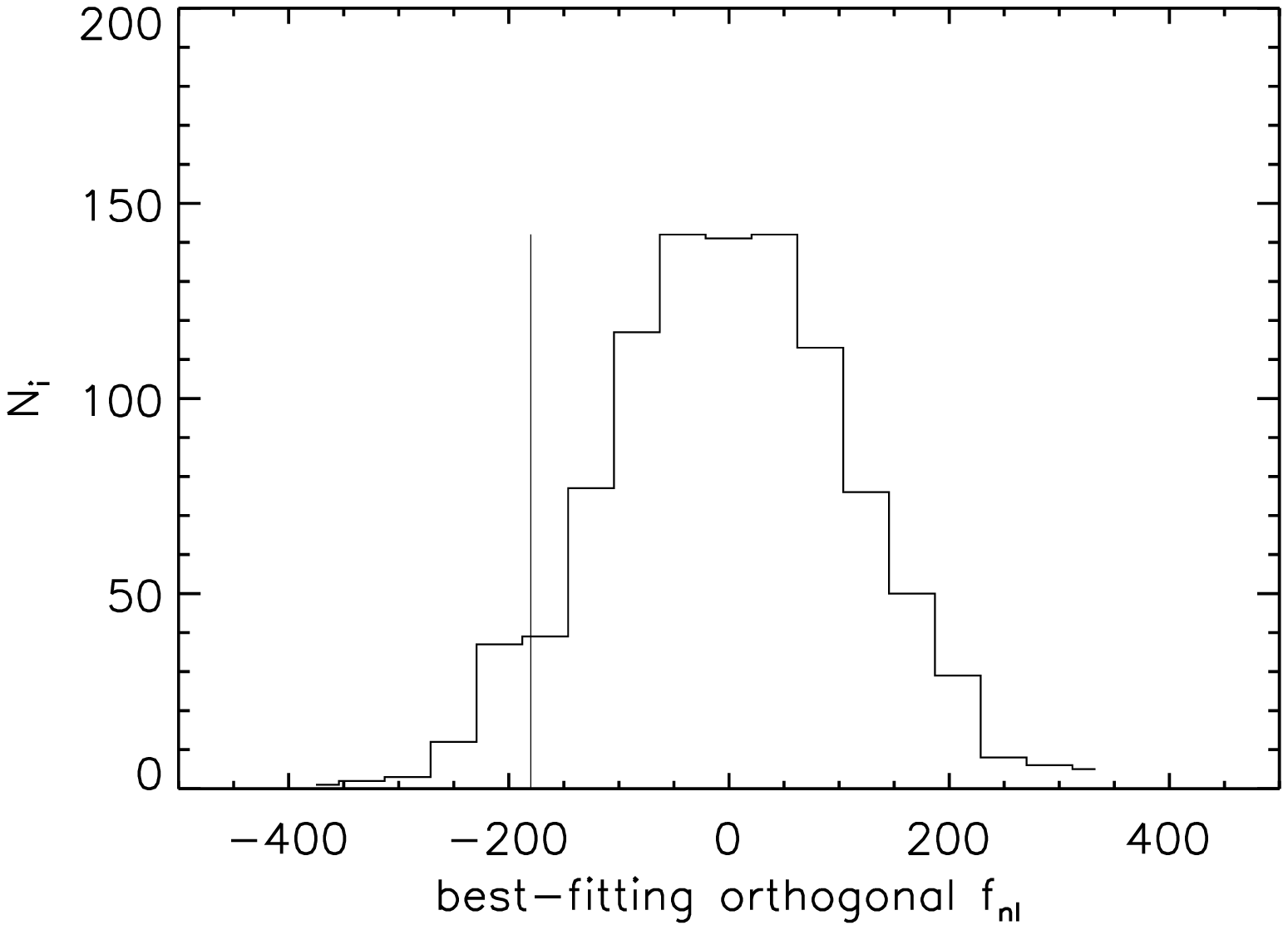}
  \caption{From left to right, the histograms with the best-fitting
  $f_{nl}$ values obtained for 1,000 Gaussian simulations for the
  local, equilateral and orthogonal shapes. Vertical lines correspond
  to the values obtained with the data after taking into account the
  point source contamination. \label{plot_bestfnl}}
\end{figure*}
Fig. \ref{plot_bestfnl} contains the histograms of the best-fitting
$f_{nl}$ values for each shape corresponding to 1,000 CMB + noise
Gaussian simulations and the values of the data after the point source
correction. Note that the point sources add a significant contribution
to the equilateral and orthogonal shapes. We agree with
\citet{komatsu2011} that the WMAP seven-year data are consistent with
Gaussian primordial fluctuations for the three considered
shapes. Planck will be able to address this issue with more detail due
to its increased sensitivity and power to clean the signal.
\section{Conclusions}
\label{conclusions}
We have imposed constraints on primordial non-Gaussianity with the
WMAP 7-year data using the wavelet based estimator. In this analysis
we have considered the combined V+W maps and the KQ75 mask. In
particular, we have focused in three shapes with particular interest
for the physics of inflation in the early universe: the local,
equilateral and orthogonal bispectra.

We have simulated the non-Gaussian maps for each of the considered
shapes and estimated with these simulations the required quantities
for our estimator. Our results are compatible with the values obtained
by the WMAP team and our uncertainties are very similar to the error
bars obtained with the optimal bispectrum estimator
\citep{komatsu2011}.

In addition we have estimated the contribution of the point
sources. In the particular case of the local $f_{nl}$, the
contribution is $\Delta f_{nl}^{loc}= 2.5 \pm 3.0$. This is
similar to the values obtained by the WMAP team
\citep{komatsu2011} and its contribution to the parameter is not
significant. However we have detected a non-negligible contribution to
the equilateral and orthogonal shapes due to the unresolved point
sources. In particular, we have found $ \Delta f_{nl}^{eq}= 37 \pm
18$ and $ \Delta f_{nl}^{ort}= 25 \pm 14$ (68\%CL). These large
values were already predicted by \citet{komatsu2011} although they did
not provide actual figures. This contribution should be taken
carefully into account in future constraints on $f_{nl}$ with WMAP and
Planck data. Considering the point sources, our best estimates of
$f_{nl}$ are $ f^{loc}_{nl}=30.0 \pm 22.5$, $ f^{eq}_{nl}=-90
\pm 146$ and $ f^{ort}_{nl}=-180 \pm 107$.  The three shapes
are compatible with zero at 95\% CL. Our conclusion is that the
$f_{nl}$ parameters are compatible with zero within the $2\sigma$ CL
and our results are in agreement with \citet{komatsu2011}.

The wavelet estimator has been tested and carefully checked with the
available WMAP data in this and several previous works
\citep{curto2009a,curto2009b,curto2011}. It is now ready and being
upgraded to analyse the forthcoming Planck data. In future works we
will also use the wavelet estimator jointly with neural networks to
constrain these shapes along the lines of \citet{casaponsa2011b} where
this procedure has been already applied for the local $f_{nl}$ using
WMAP data. This later process helps to speed up the calculations since
it is not necessary to estimate the covariance matrix of the cubic
statistics and it avoids all the possible complications in the
computation of the inverse covariance matrix.
\section*{acknowledgments}
The authors are thankful to Eiichiro Komatsu and Michele Liguori for
their useful comments that have helped in the production of this
paper. The authors thank J. Gonz\'alez-Nuevo for providing the $dN/dS$
counts and for his useful comments on unresolved point sources. The
authors also thank Biuse Casaponsa, Airam Marcos-Caballero, Sabino
Matarrese and Patricio Vielva for useful comments on different
computational and theoretical issues on the primordial
non-Gaussianity. The authors acknowledge partial financial support
from the Spanish Ministerio de Ciencia e Innovaci\'on project
AYA2010-21766-C03-01, the CSIC-Royal Society joint project with
reference 2008GB0012 and the Consolider Ingenio-2010 Programme project
CSD2010-00064. A. C. thanks the Universidad de Cantabria for a
post-doctoral fellowship. The authors acknowledge the computer
resources, technical expertise and assistance provided by the Spanish
Supercomputing Network (RES) node at Universidad de Cantabria. We
acknowledge the use of Legacy Archive for Microwave Background Data
Analysis (LAMBDA). Support for it is provided by the NASA Office of
Space Science. The HEALPix package was used throughout the data
analysis \citep{healpix}.
%
%

%
\end{document}